\documentclass{article}
\usepackage{graphicx}   % Required for inserting images

\usepackage[utf8]{inputenc}
\usepackage[T1]{fontenc}

\usepackage{amsmath}
\usepackage{bm}         % bold math
\usepackage{siunitx}    % For correct automatic typesetting of units
\usepackage{authblk}    % Author affiliation
\usepackage{url}        % For bibliography
\usepackage{float}      % Float environment

\usepackage{newpxtext, newpxmath}   % Palatino
\usepackage[dvipsnames]{xcolor}
\usepackage[normalem]{ulem}  % Strike out
\usepackage[margin=38mm]{geometry}  % Slightly more sensible margins to make equation fit
\usepackage{todonotes}

\newcommand{\abs}[1]{\left|#1\right|}
\newcommand{\abssq}[1]{\abs{#1}^2}
\newcommand{\conj}[1]{{#1}^*}
\newcommand{\inner}[2]{\left\langle #1, #2\right\rangle}
\DeclareMathOperator*{\argmin}{arg\,min} % thin space, limits underneath in displays

\newcommand{\opl}{\mathrm{OPL}}  % Optical Path Length
\newcommand{\zernike}{Z}  % Zernike symbol
\newcommand{\zerncf}[1]{a_{#1}}  % Zernike mode coefficient

\newcommand{\rwaist}{r_\mathrm{waist}}  % Gaussian waist radius
\newcommand{\NA}{\mathrm{NA}}  % Numerical Aperture
\newcommand{\ampxy}{A(x, y)}  % Beam amplitude profile (field)
\newcommand{\amp}{A}
\newcommand{\cfx}[3]{\alpha_{#1#2#3}}
\newcommand{\cfy}[3]{\beta_{#1#2#3}}
\newcommand{\transf}[1]{#1'} 
\newcommand{\phasexy}[1]{\phi_{#1}(x, y)}  % Phase function
\newcommand{\wphasexy}[1]{\phi_{#1}(\transf{x}, \transf{y})}  % Warped phase function
\newcommand{\fun}{E}  % Basis function
\newcommand{\wfun}{E'}  % Transformed basis function

\newcommand{\nonortho}{\mathcal{N}}  % Non-orthogonality
\newcommand{\meanphasegrad}{\mathcal{G}}  % Phase grad
\newcommand{\im}{m}     % Mode index 1
\newcommand{\iM}{M}     % Number of modes
\newcommand{\imm}{n}    % Mode index 2
\newcommand{\ix}{p}     % Polynom exponent/index x
\newcommand{\iy}{q}     % Polynom exponent/index y
\newcommand{\iz}{j}     % Zernike index

\newcommand{\kx}{k_x^{(\im)}}
\newcommand{\ky}{k_y^{(\im)}}

\newcommand{\perfimprovement}{1.5 } %%%% TODO: update when new measurements are done
\newcommand{\nummeasurements}{35 }

\title{Orthonormalization of phase-only basis functions}
\author{D.W.S. Cox and I.M. Vellekoop}
\date{July 2024}

\begin{document}

\maketitle
\hrule

\begin{abstract}
    % General context, i.e. this stuff very useful :)
    Orthonormal bases serve as a powerful mathematical tool in theoretical and experimental optics.
    % Problem context
    However, producing arbitrary optical fields in real-world experiments is limited by the hardware, which in many cases involves a phase-only spatial light modulator.
    % Problem statement
    Since most basis functions also have a varying amplitude component, they cannot be represented truthfully.
    % Our contribution
    We present a general method to construct an orthonormal phase-only basis, optionally possessing desirable properties like smoothness and symmetry.
    % Demo that it's not just awesome, but also useful
    We demonstrate the practical benefit of our approach in a wavefront shaping experiment, achieving a factor \perfimprovement increase in performance over a non-orthonormal phase only basis.
\end{abstract}

\section{Introduction}

\label{sec:intro}

% Modes als oplossing van een differentiaalvergelijking
Orthonormal basis functions are an indispensable concept in many fields of research and engineering \cite{Friedberg2003LinearAlgebra}. In optics, orthonormal basis functions commonly arise as solutions to wave equations. The archetypal example are the plane waves, which are solutions to the free space Maxwell's equations \cite{Dandliker1999ConceptModes} and form the basis of the angular spectrum method \cite{goodmanIntroductionFourierOptics2005}.
Other common examples are Laguerre-Gauss and Hermite-Gauss bases, which are solutions to the paraxial wave equation, and the TE, TM and TEM bases used to analyze wave propagation in a waveguide \cite{Saleh1991FundamentalsPhotonics}.
The research field of structured light has recently seen interesting advancements in the generation of different types of orthonormal basis functions. However, many of these types of modes require complex modulation to control both amplitude and phase.
% \cite{Rosales-GuzmanCompactCookbook} ?  %%% TODO

% In optical experiments, arbitrary optical fields can be constructed using a variety of devices, including pixelated spatial light modulators (SLMs), deformable mirrors and diffractive optical elements \cite{Gutierrez-Cuevas2023BinaryHolograms}.
% A widely used
% As many of these devices only allow the phase of the light to be modulated, it is natural to ask whether there are phase-only orthonormal bases, for which all basis functions have the same amplitude profile.
A widely used device to construct arbitrary optical fields is the phase-only spatial light modulator (SLM).
Hence, a natural choice of an orthonormal basis is a phase-only basis, as these can be trivially displayed on a phase-only SLM.
Unfortunately, there are currently very few options to choose such a basis. Known examples are the Hadamard basis \cite{popoff2010MeasuringTransmission} and the plane-wave basis \cite{mastianiWavefrontShapingForward2022}, which are orthonormal on a homogeneously illuminated rectangle, or those bases mapped to a disk \cite{yap2010TwoDimensionalPolar,mastiani2021NoisetolerantWavefront}.  

Here we describe a general method for constructing an orthonormal basis from an arbitrary set of phase patterns. Our method finds a near-orthonormal basis for any specified illumination profile, while preserving desirable properties like smoothness and symmetry. We demonstrate experimentally that this new basis can be used as a drop-in replacement in a wavefront shaping experiment, achieving a factor of \perfimprovement higher improvement in the signal intensity.

% Onze bijdrage
This article is composed of three parts. In the first part, we lay down the theoretical groundwork of our method. In the second part, we apply our method to orthonormalize a plane wave basis, and demonstrate the effectiveness of this basis in a wavefront shaping experiment. In the last part, we illustrate that our method is generally applicable with a second example: an orthonormal basis based on Zernike modes. The Python code to produce these results is available on GitHub \cite{Dedean16Phase_only_orthonormalization}.

\section{Theory}
\label{sec:theory}

\subsection{Orthonormality condition}
\label{subsec:inner-product}
% Assumptions and inner product definition
We consider the situation where a phase-only SLM is illuminated with a linearly polarized beam with field amplitude $A(x,y)$, with spatial coordinates $x$ and $y$. We have direct control over the phase pattern $\phasexy{}$ on the SLM. Our goal is to construct a smooth, orthonormal basis by only varying the phase patterns, based on an existing basis. Additionally, we wish to maintain $x,y$-symmetries in the basis functions, if any.

Each basis function $\fun_{\im}$ is given by: $E_m(x,y)=A(x,y)\,\exp{[i\phi_m(x,y)]}$, where $\im$ is the index of the basis function.
The inner product of two functions $\fun_\im$ and $\fun_\imm$ is defined as:
\begin{equation}\label{eq:inner-product-int}
\begin{split}
    \inner{\fun_{\im}}{\fun_{\imm}}
    & = \iint_{-\infty}^{\infty}
        \conj{\fun{}}_\im(x, y) \,\fun_{\imm}(x, y)
        \;dx\,dy\\
    & = \iint_{-\infty}^{\infty}
        \amp^2(x,y)\,e^{i(\phi_\imm(x,y)-\phi_\im(x,y))} 
        \;dx\,dy
\end{split}
\end{equation}
where $\conj{}$ denotes complex conjugation. The beam amplitude $\amp$ is normalized such that $\inner{\fun_{\im}}{\fun_{\im}} = 1$.
The inner products of all combinations of $m$ and $n$ form the Gram matrix $G$:
\begin{equation}\label{eq:gram-matrix}
    G_{\im\imm}=\inner{\fun_{\im}}{\fun_{\imm}}
\end{equation}
For an orthonormal basis, $G$ is equal to the identity matrix $I$. Thus, our goal is to find a basis $\{\fun_{\im}\}$ with $G$ as close to $I$ as possible.

\subsection{Method}
\label{subsec:method}
To construct an orthonormal basis, we start with an initial set of smooth phase patterns and a global beam amplitude profile, which together form a non-orthonormal basis.
\begin{equation}\label{eq:basis-original}
    \fun_{\im}(x, y) = \ampxy \, e^{i \phasexy{\im}}
\end{equation}

The goal is to find an orthonormal basis that somehow maintains smoothness and symmetry of the initial phase patterns $\phasexy{\im}$, while keeping the amplitude profile equal for each basis function. We do so by applying a smooth coordinate transform $x,y\rightarrow \transf{x}_{\im},\transf{y}_{\im}$ to each phase pattern individually, to produce basis functions:
\begin{equation}\label{eq:warped-tilt}
    \wfun_{\im}(x, y) = \ampxy \, e^{i \phi_{\im}(\transf{x}_{\im},\transf{y}_{\im})}
\end{equation}

The new basis $\{\wfun_{\im}\}$ should be as close to orthonormal as possible, while preserving smoothness. We define the coordinate transformation as two bivariate polynomials:
\begin{equation}\label{eq:xwarped}
    \transf{x_\im}(x, y) = x \!\sum_{\ix=0}^P\sum_{\iy=0}^P \cfx{\ix}{\iy}{\im} x^{\ix} y^{\iy}
\end{equation}
\begin{equation}\label{eq:ywarped}
    \transf{y_\im}(x, y) = y \!\sum_{\ix=0}^P \sum_{\iy=0}^P \cfy{\ix}{\iy}{\im} x^{\ix} y^{\iy}
\end{equation}
where $x, y$ are the original coordinates, and $\ix$ and $\iy$ are non-negative integers in the range $[0, P]$. $P$ is manually chosen to give the transformation sufficient degrees of freedom, while keeping the transform smooth.
$\cfx{\ix}{\iy}{\im}$ denotes the polynomial coefficients for the $\transf{x}$ transformation and $\cfy{\ix}{\iy}{\im}$ denotes the polynomial coefficients for the $\transf{y}$ transformation.
Note that both polynomials contain no offset coefficient, i.e.: $\transf{x}(0, 0) = 0$ and $\transf{y}(0, 0)=0$.
If desired, spatial symmetries in $x$ and $y$ can be preserved by restricting to even numbers $\ix$ and $\iy$.

Our task now is to find coefficients $\cfx{\ix}{\iy}{\im}$, $\cfy{\ix}{\iy}{\im}$ that maximize the orthonormality. We quantify a metric for \emph{non-orthonormality} $\nonortho$ as a normalized squared Frobenius norm:
\begin{equation}\label{eq:non-orthonormality}
    \nonortho = \frac{1}{\iM} \sum_{\im} \sum_{\imm} \abssq{G_{\im \imm} - I_{\im \imm}}
\end{equation}

where $\iM$ is the number of modes.
A perfectly orthonormal basis has $G=I$, resulting in the minimum value of $\nonortho=0$.

One solution to minimize the non-orthonormality $\nonortho$ is with polynomial transforms with very large coefficients $\cfx{\ix}{\iy}{\im}, \cfy{\ix}{\iy}{\im}$, which causes very steep phase gradients.
Though highly orthonormal, this is an undesirable solution, as we want to find smooth phase functions. We avoid this unwanted solution by regularizing the problem.
We define a metric for the phase gradient magnitude:
\begin{equation}\label{eq:mean-phase-gradient}
    \meanphasegrad = 
    \frac{1}{M}\sum_{\im}{
        \iint_{-\infty}^{\infty} \amp^2(x,y)
        \abssq{\nabla \wphasexy{\im}} \;dx\,dy
    }
\end{equation}
where $\nabla$ denotes the gradient with respect to the coordinates $x,y$.

We combine the above two definitions into the following minimization problem:

\begin{equation}\label{eq:minimization-short}
    \argmin_{\text{all } \cfx{\ix}{\iy}{\im},\,\cfy{\ix}{\iy}{\im}} \;
    \nonortho + \frac{1}{w^2} \meanphasegrad
\end{equation}
where $w$ is a meta-parameter that represents a tolerable phase gradient in \si{rad / pixel}.
The first term helps to improve the orthonormality and the second term helps to find a solution with low phase gradients.
The weight $w$ can be tuned to set the importance of finding a smooth solution. 
Lower values for $w$ will typically result in better orthonormality, but this can be at the cost of steeper phase gradients. Higher values for $w$ will typically result in smaller phase gradients, but with worse orthonormality.
The optimal balance between these two aspects and the corresponding ideal value of $w$, will depend on the application at hand.

We solve the minimization problem of Eq.~\ref{eq:minimization-short} with a the non-convex optimization method AMSGrad \cite{reddiConvergenceAdam2018}.
We initialize the coefficients to $\cfx{0}{0}{\im} = \cfy{0}{0}{\im} = 1$ for all $\im$, and all other elements to 0, such that initially $\wfun=\fun$.
Our implementation can be found at \cite{Dedean16Phase_only_orthonormalization}.

\section{Orthonormalizing plane waves}
\label{sec:results-discuss}

% Onze demo-applicatie
In this section, we demonstrate the construction of a smooth phase-only orthonormal basis with plane waves as initial basis. We will show that the orthonormalized basis can be used as a drop-in replacement in a wavefront shaping experiment.

Wavefront shaping is an technique to focus light through scattering media \cite{kubbyWavefrontShapingBiomedical2019}.
Although wavefront shaping works with any basis \cite{vellekoopFocusingCoherentLight2007a}, an orthonormal basis is optimal (please see section 2 of the supplementary materials for more information).
Many different approaches exist to compose a wavefront shaping algorithm \cite{gigan2022RoadmapWavefront}.
A common approach is to split the SLM into a part that displays the basis functions, and a part that acts as a reference \cite{mastianiWavefrontShapingForward2022, tao2017ThreedimensionalFocusing, popoff2010MeasuringTransmission}.
% The contribution of each basis function to the field in the target focus (i.e. a transmission matrix element) is then typically measured with phase stepping interferometry. From this information, the optimal wavefront that focuses to the target can be computed.
%%%%% TODO? Our WFS algorithm uses..?
Please see section 4 of the supplementary materials for more details of our measurement procedure.

To construct an optimal basis for this experiment, we first consider the amplitude profile at the SLM. Our SLM is illuminated with a Gaussian beam, and imaged onto the back-pupil of the microscope objective (see supplementary section 1 for a description of the experimental setup). The objective's pupil aperture truncates the Gaussian profile, effectively creating a truncated Gaussian illumination profile, which is often used in microscopy \cite{Marshall2012HandbookOptical}.
In our wavefront shaping algorithm, the SLM active area is divided in two halves, with one half being used to display the basis functions. Therefore, we are looking for basis functions that are orthonormal on half of the SLM with the amplitude profile:
\begin{equation}
    \label{eq:gauss-beam-half}
    \ampxy = 
    \begin{cases}
        \amp_0\exp{\left[ -\frac{r^2}{\rwaist^2} \right]}
            &\text{if }r\leq 1 \text{ and } x<0\\
        0  & \text{if }r>1 \text{ or } x\geq 0
    \end{cases}
\end{equation}
where $r=\sqrt{x^2 + y^2}$, and $\rwaist$ is the waist of the Gaussian illumination.
$x$, $y$ and $\rwaist$ are normalized such that the objective's pupil aperture is at $r=1$. $\amp_0$ denotes the normalization factor such that $\inner{\amp}{\amp}=1$.

In this example, we will use an improved version of the wavefront shaping algorithm by Tao~et~al. \cite{tao2017ThreedimensionalFocusing}. Originally, this algorithm uses a Hadamard basis with small square segments. It was already shown that a plane wave basis is more efficient even with strongly scattering samples \cite{mastianiWavefrontShapingForward2022}.
However, neither the Hadamard nor the plane wave basis is orthonormal for a truncated Gaussian amplitude profile. Therefore, we expect to reach higher signal enhancements using our orthonormalized basis, without modifying the number of measurements or any other aspect of the wavefront shaping algorithm. (In section 2 of the supplementary materials we explain why we expect better performance with an orthonormal basis.)

\subsection{Constructing the orthonormal basis}\label{subsec:construct-ortho-plane-wave-basis}

As an initial basis we choose plane wave functions that span half of the pupil in $x$ and the whole pupil in $y$, as used in Mastiani~et~al. \cite{mastianiWavefrontShapingForward2022}:
\begin{equation}
    \phasexy{\im} = \left(\kx x + \ky y\right)
\end{equation}
where $x,y$ are normalized coordinates such that the objective's pupil aperture is at $r=1$, $\kx$ are integer multiples of $\pi$ and $\ky$ are integer multiples of $\pi/2$.
These values make $e^{i\phasexy{\im}}$ orthonormal on the rectangular domain $x\in[-1, 0]$ and $y\in[-1, 1]$. Correspondingly, a phase gradient with $\ky=\pi/2$ shifts the focus in the focal plane by exactly one diffraction limit $\lambda/(2\,\NA)$. For $\kx$, only half the pupil is used, so the diffraction limit is doubled w.r.t. $\ky$.
% ===== Derivation =====
% k = 2π/λ                                      % k = wavenumber = angular spatial frequency
% r = pupil radius = f⋅n⋅sin(θ_NA) = f⋅NA       % Abbe sine condition
% Δk = 2π/(2r) = π/r = π/(f⋅NA)                 % Angular base frequency for k_y
% Δx = f⋅Δk/k                                   % Fourier optics
%    = f⋅Δk⋅λ/2π = f⋅π⋅λ / (2π⋅f⋅NA) = λ/(2⋅NA) = diffraction limit
% ======================

\begin{figure}
    \centering
    \includegraphics[width=1.0\linewidth]{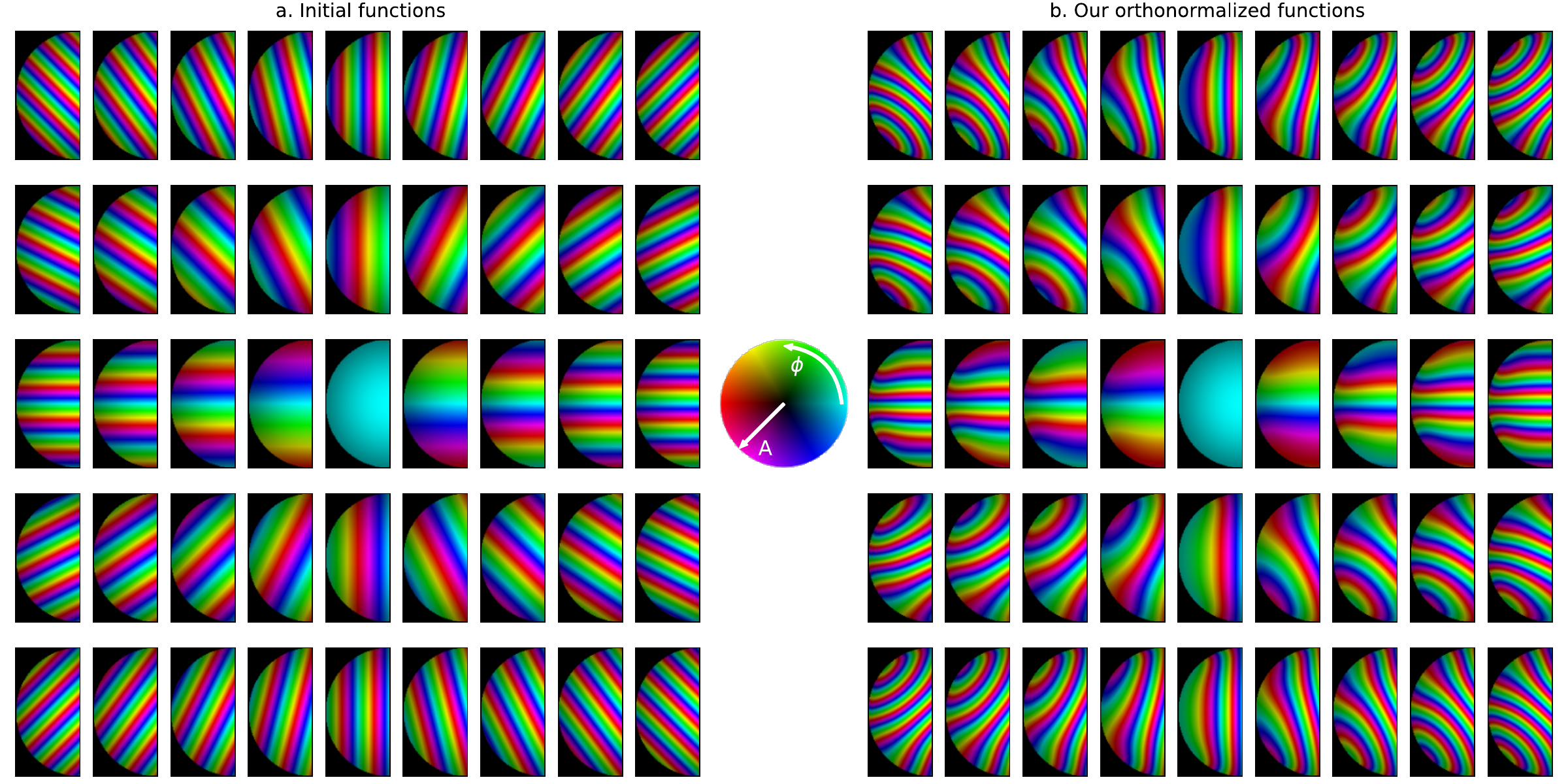}
    \caption{a. Initial plane wave basis. b. Our new orthonormal basis. Each plot shows a basis function in the domain $x\in[-1, 0]$ (horizontal) and $y\in[-1, 1]$ (vertical). The colors illustrate the complex valued functions, where the hue represents the phase $\phi$ and the brightness represents the amplitude $\amp$. Black represents $0$ and the maximum brightness represents the maximum amplitude value. Both bases consist of smooth functions with relatively low phase gradients. Moreover, the new basis functions approximately preserve the phase gradient magnitude and direction of the initial basis functions.}
    \label{fig:init-new-plane-waves}
\end{figure}

\begin{figure}
    \centering
    \includegraphics[width=1.0\linewidth]{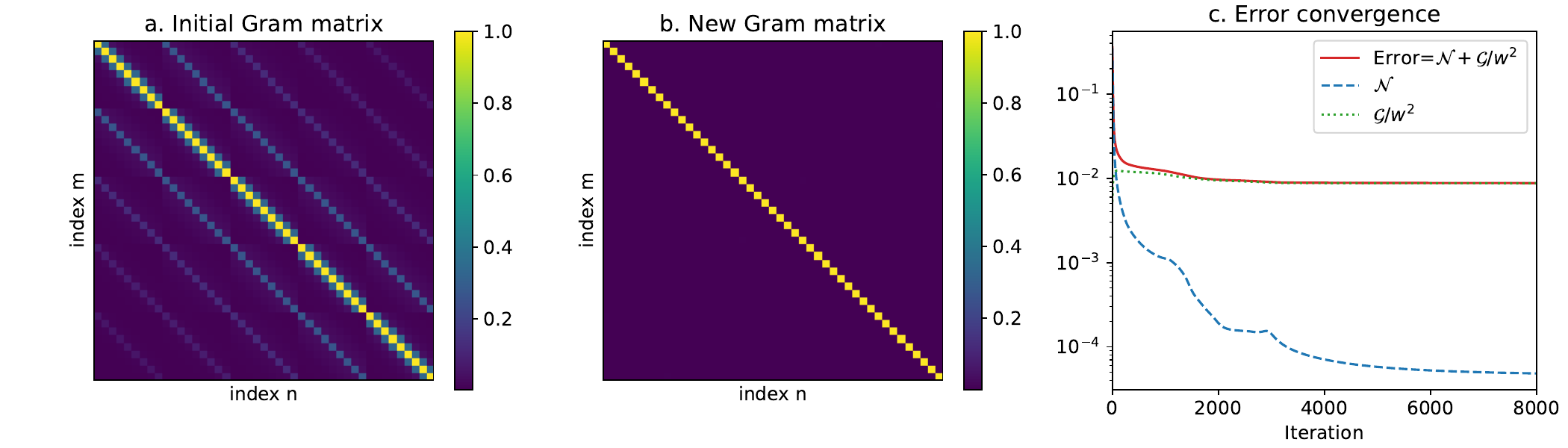}
    \caption{a. Gram matrix of the initial plane wave basis. b. Gram matrix of our new orthonormal basis. The color plot shows the absolute value of each matrix element. Non-zero off-diagonal elements indicate non-orthonormality. c. Convergence of minimizing the error function and its terms: non-orthonormality $\nonortho$ and the phase gradient magnitude term $\meanphasegrad/w^2$.}
    %%% TODO: update figure
    \label{fig:plane-waves-gram-error}
\end{figure}

We limit $\{\phasexy{\im}\}$ to the ranges $\kx\in[-4\pi, 4\pi]$ and $\ky\in[-4\pi, 4\pi]$, resulting in a total of 45 basis functions. 
Fig.~\ref{fig:init-new-plane-waves}a shows the functions of this initial basis. With a non-orthonormality $\nonortho=0.35$, this basis is clearly not orthonormal. This fact can also be concluded by observing the non-zero off-diagonal elements of the Gram matrix in Fig.~\ref{fig:plane-waves-gram-error}a.

We initialize the coefficients to $\cfx{0}{0}{\im} = \cfy{0}{0}{\im} = 1$ and all other elements to 0, such that initially $\wfun=\fun$.
Next, we minimize the error function of Eq.~\ref{eq:minimization-short} with AMSGrad optimization \cite{reddiConvergenceAdam2018} in $8000$ iterations. We set the phase gradient weight $w$ to \SI{2.2}{rad/pix}, and choose even powers $\iy$ to preserve the evenness/oddness of the functions in $y$.
We found that the ranges $\ix=\{0,1,2,3,4,5,6,7\}$ and $\iy=\{0,2,4,6,8,10\}$
were sufficient to reach a non-orthonormality of $\nonortho=0.00005$.
Fig.~\ref{fig:plane-waves-gram-error}c shows the convergence of the error function. The error function has converged to a minimum. Fig.~\ref{fig:init-new-plane-waves}b shows the new basis functions.
As can be observed, the new basis functions preserve the smoothness and approximate phase gradient magnitude and direction.
Most of all, with $\nonortho=0.00005$, the new basis is almost perfectly orthonormal. This can also be observed from the new Gram matrix $G$ in Fig.~\ref{fig:plane-waves-gram-error}b, which is an almost perfect identity matrix.

\subsection{Improved wavefront shaping}
\label{subsec:experiment-wfs}

\begin{figure}
    \centering
    \includegraphics[width=0.98\linewidth]{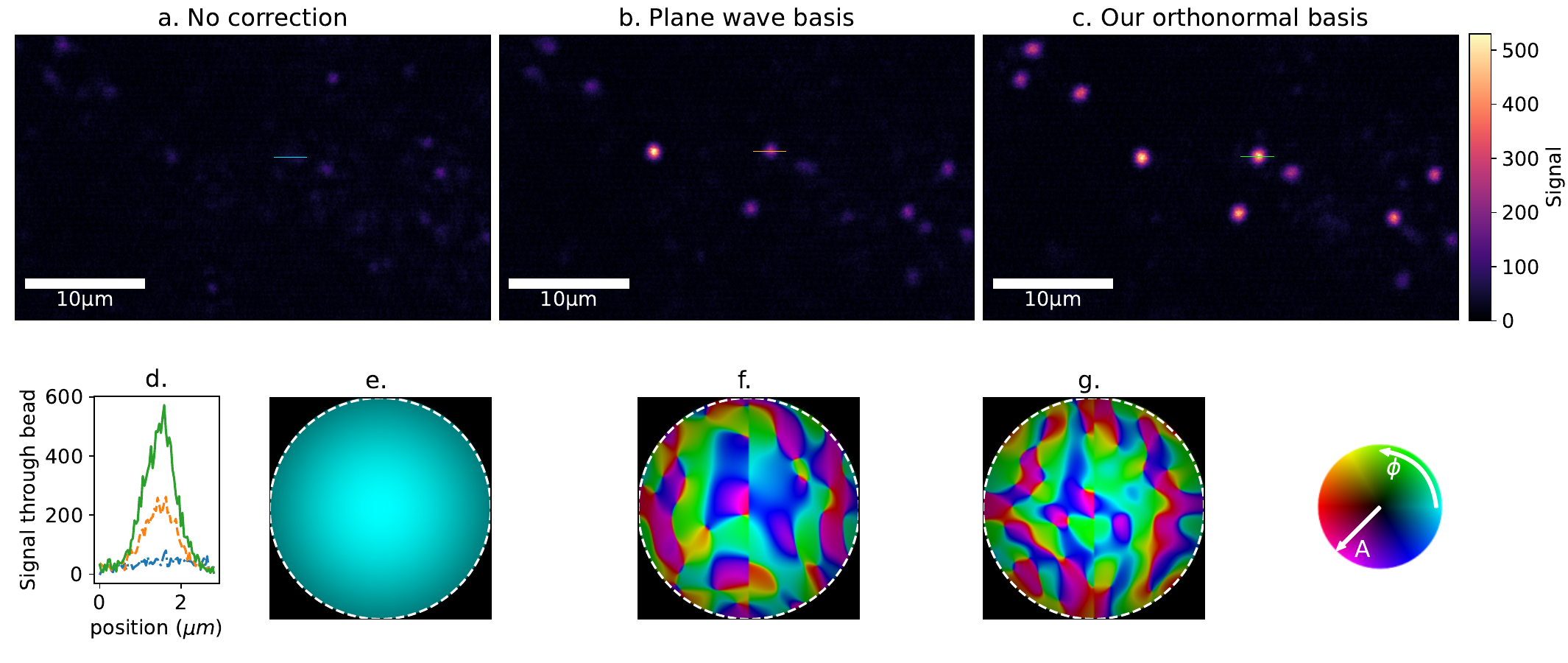}
    \caption{(a, b, c) Images of \SI{500}{nm} fluorescent beads, \SI{250}{\micro m} beneath a scattering PDMS surface. (a) without correction pattern. (b) correction using a plane wave basis. (c) correction with our orthonormal basis.
    (a, b, c) show line segments through one of the beads.
    The signal intensity along these line segments is plotted in (d) for the uncorrected image (dash-dotted blue), the image corrected with a plane wave basis (dashed orange) and our orthonormal basis (solid green).
    (e, f, g) show the pupil fields from the illuminated SLM for recording image (a, b, c) respectively.
    The white dashed line indicates the pupil aperture.}
    \label{fig:images-beads}
\end{figure}

\begin{figure}
    \centering
    \includegraphics[width=0.6\linewidth]{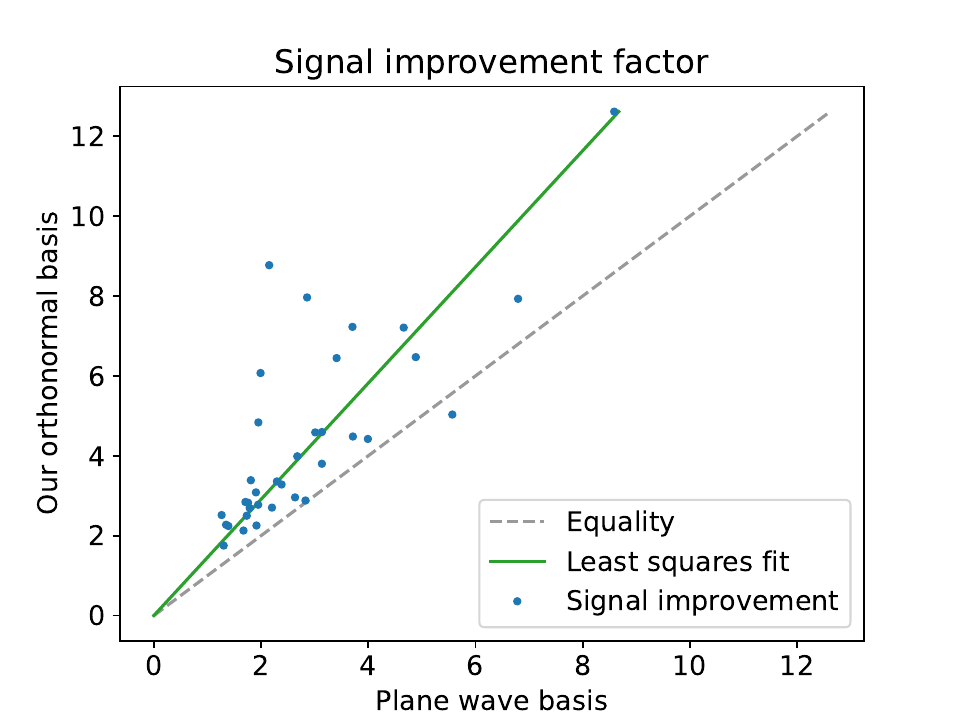}
    \caption{Signal improvement factor achieved by wavefront shaping with the original plane wave basis vs our orthonormalized plane wave basis. Both bases have 45 basis functions. By switching to our basis, the signal improvement factor is increased by a factor \perfimprovement on average (least squares fit).
    }
    \label{fig:wfs-plane-wave-vs-ortho}
\end{figure}

We now apply our newly constructed orthonormal basis in a wavefront shaping experiment. Our setup is a 2-photon excitation fluorescence (2PEF) laser-scanning wavefront shaping microscope. A schematic of the setup can be found in the supplementary materials.
Our sample is a slab of PDMS with a scattering surface and dispersed \SI{500}{nm} fluorescent beads inside.
The goal of our experiment is to improve the focus, and thus the imaging quality, beneath the scattering surface of the sample.

Fig.~\ref{fig:images-beads}a shows an image of the fluorescent beads, \SI{250}{\micro m} beneath the scattering surface. The beads appear very faintly. We optimize the focus with the wavefront shaping algorithm.
Due to the non-linear 2PEF-feedback, a sharper focus will increase the feedback signal intensity and signal to noise ratio \cite{Osnabrugge2019BlindFocusing, Katz2014NoninvasiveNonlinear}. 
Fig.~\ref{fig:images-beads}bc show the same beads as Fig.~\ref{fig:images-beads}a, but were imaged while displaying a phase correction pattern on the SLM. Both corrected images are significantly brighter than the uncorrected image. Furthermore, the image obtained with our orthonormal basis (Fig.~\ref{fig:images-beads}c) is significantly brighter than the image obtained with the original plane wave basis (Fig.~\ref{fig:images-beads}b). Fig.~\ref{fig:images-beads}a,b,c each show a line segment through one of the beads. The signal intensity along these line segments is plotted in Fig.~\ref{fig:images-beads}d for the uncorrected image (dash-dotted blue), the image corrected with a plane wave basis (dashed orange) and our orthonormal basis (solid green). By switching to our orthonormal basis, this bead appears over twice as bright.

Fig.~\ref{fig:images-beads}e,f,g show the field produced by the illuminated SLM when imaging Fig.~\ref{fig:images-beads}a,b,c respectively.
In these color plots, amplitude is shown as brightness and the phase is shown as hue.
Note that our SLM is a phase-only device. Therefore, we use only the phase of the measured correction to display on the SLM. On average, the lack of amplitude modulation results in a factor of $\pi/4$ lower enhancement \cite{Mastiani2024PracticalConsiderationsb}.

The degrees of freedom from the plane wave basis cover a rectangular domain.
The degrees of freedom from our orthonormal basis cover a smaller circular domain (i.e. the pupil).
Therefore, we expect the orthonormalized plane wave basis to be able represent higher spatial frequencies, and thus more details, inside the circular domain.
In agreement, we see that the correction pattern measured with our orthonormal basis clearly has more details (Fig.~\ref{fig:images-beads}g) than the pattern measured with the plane wave basis (Fig.~\ref{fig:images-beads}f).
Essentially, using the plane wave basis, which is orthonormal on a square domain, wastes measurements on reconstructing the amplitude profile, including unilluminated sections outside the pupil. By contrast, no measurements are `spent' on the unilluminated corners when using our orthonormal basis, leaving more degrees of freedom for the illuminated pupil.

We have repeated the wavefront shaping measurements on \nummeasurements
different locations on the same sample. At each location, we compared the performance of the original plane wave basis with that of our new orthonormalized plane wave basis. This comparison is shown in Fig.~\ref{fig:wfs-plane-wave-vs-ortho}. For every location, our orthonormalized basis performed better than the original plane wave basis. Our orthonormalized basis increased the average signal improvement factor by a factor \perfimprovement (least squares fit).

\section{Orthonormalizing fields from Zernike polynomials}
\label{sec:ortho-zernike}

In this section, we demonstrate that our method can be applied more generally by constructing an orthonormal basis starting from Zernike polynomials.
Zernike polynomials are often used in adaptive optics (AO) \cite{Beckers1993AdaptiveOptics, kubbyAdaptiveOpticsBiological2013} for smooth aberration correction.
Zernike polynomials are designed to be orthonormal in the optical path length ($\opl$). However, the corresponding fields are proportional to $e^{i k \opl}$ and thus are \emph{not} orthonormal. In the following section, we will optimize fields from Zernike polynomials for wavefront shaping with our orthonormalization method.

We will use the following truncated Gaussian amplitude function $\ampxy$:
\begin{equation}
    \label{eq:gauss-beam}
    \ampxy = 
    \begin{cases}
        \amp_0\exp{\left[ -\frac{r^2}{\rwaist^2} \right]}  &\text{if }r\leq 1 \\
        0  & \text{if }r>1
    \end{cases}
\end{equation}
This amplitude function is similar to the amplitude function of the previous section, except that it spans the entire objective pupil instead of half the pupil. This makes it suitable for wavefront shaping with an external reference field.

\begin{figure}
    \centering
    \includegraphics[width=1\linewidth]{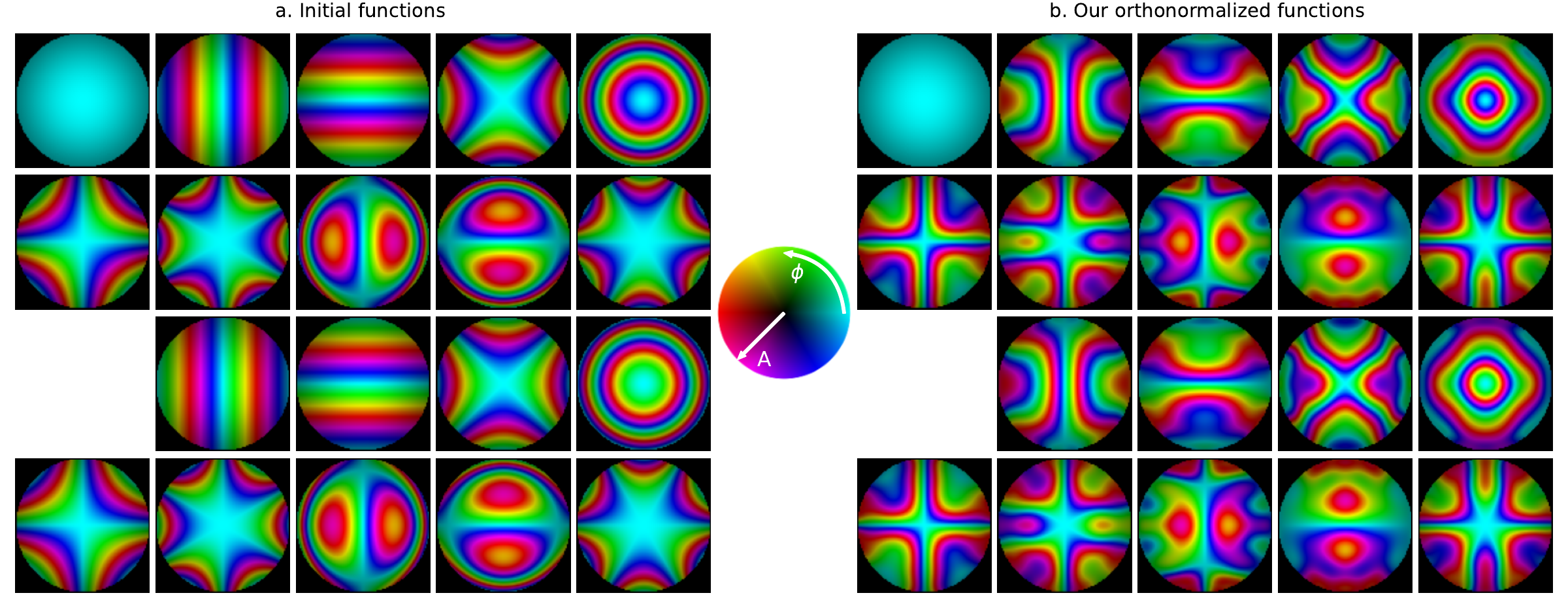}
    \caption{\label{fig:init-and-new-zernike}
    a. Initial basis of fields from Zernike polynomials. b. A newly constructed orthonormal basis. Each plot shows a basis function in the domain $x\in[-1, 1]$ (horizontal) and $y\in[-1, 1]$ (vertical). The colors illustrate the complex valued functions, where the hue represents the phase $\phi$ and the brightness represents the amplitude $\amp$. Just like the initial basis, the new basis functions are smooth, have small phase gradients, and retain their evenness and oddness in $x$ and $y$.}
\end{figure}

Next, we choose the phase functions $\phasexy{\im}$ of our initial basis as produced with the first 10 Zernike polynomials:
\begin{equation}
    \phasexy{\im} = \zerncf{\im} \zernike_{\iz} (x,y)
\end{equation}

where $\zernike_\iz$ indicates the $\iz$th Zernike polynomial and $\zerncf{\im}$ denotes the Zernike coefficient, which is a real constant.
For our initial basis, we pick each Zernike polynomial with $\zerncf{\im}=2\pi$ and $\zerncf{\im}=-2\pi$, except for the piston function $\zernike_1(x,y)=1$, which we include only once.
The resulting initial basis functions $\fun_{\im}$ are shown in Fig.~\ref{fig:init-and-new-zernike}a.

Secondly, we want to preserve the spatial symmetries in $x$ and $y$. Thus, we restrict Eq.~\ref{eq:xwarped} and Eq.~\ref{eq:ywarped} to even powers only: $\ix=0,2,4,6,8,10$ and $\iy=0,2,4,6,8,10$.

Thirdly, must choose a value for the phase gradient weight $w$. We would like to find a basis with orders of magnitude lower non-orthonormality (as quantified by Eq.~\ref{eq:non-orthonormality}). Additionally, we would like to preserve smoothness. With some trial and error, we find that $w=\SI{0.7}{rad/pixel}$ gives a reasonable trade-off between non-orthonormality and phase gradient.

\begin{figure}
    \centering
    \includegraphics[width=1\linewidth]{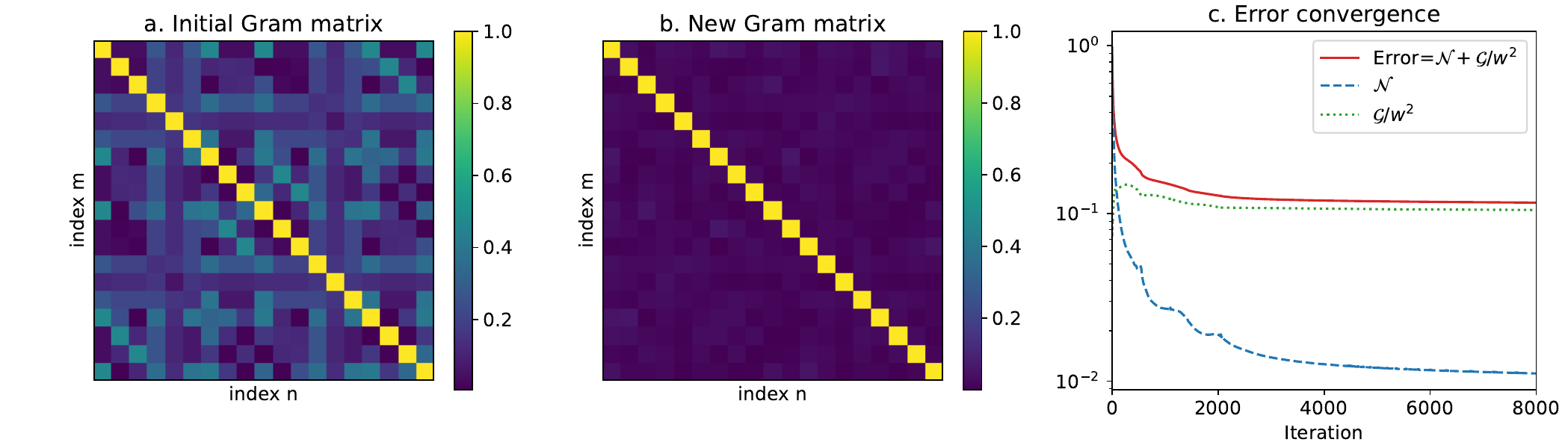}
    \caption{a. Gram matrix of the initial basis of fields from Zernike polynomials. Gram matrix the new orthonormal basis. The color plot indicates the absolute value of each matrix element. Non-zero off-diagonal elements indicate non-orthonormality. c. Convergence of minimizing the error function and its terms: non-orthonormality $\nonortho$ and the phase gradient magnitude term $\meanphasegrad / w^2$.}
    \label{fig:zernike-gram-error}
\end{figure}

The initial basis functions can be observed in Fig.~\ref{fig:init-and-new-zernike}a. Fig. \ref{fig:zernike-gram-error}a shows the absolute values of the corresponding Gram matrix. Most off-diagonal elements are significantly larger than 0, indicating that this initial basis is not orthonormal. Correspondingly, the non-orthonormality $\nonortho$ of this initial basis is \num{0.89}.
%%% Check figures

We minimize Eq.~\ref{eq:minimization-short} with AMSGrad optimization in \num{8000} iterations. Fig.~\ref{fig:init-and-new-zernike}b shows the new basis functions.
As can be observed, the new basis functions preserve the smoothness and evenness/oddness in $x$ and $y$, as can also be seen in the original Zernike modes.
Furthermore, from Fig.~\ref{fig:init-and-new-zernike}b we see that the new basis functions have relatively small phase gradients.
Most of all, they are highly orthonormal. This last fact can be observed from the new Gram matrix in Fig.~\ref{fig:zernike-gram-error}b, which closely resembles an identity matrix. Our method managed to lower the non-orthonormality $\nonortho$ to \num{0.011} for the new basis.

\section{Discussion and conclusion}
\label{sec:discussion-conclusion}

% Limitations
We have demonstrated a general method that can produce smooth orthonormal phase-only bases for arbitrary amplitude profiles.
We demonstrated our method using two different initial bases: a plane wave basis and a basis constructed from Zernike polynomials.

We have shown that the orthonormalized plane wave basis can be used as drop-in replacement in existing wavefront shaping algorithms.
This replacement yields a significant performance gain: a factor \perfimprovement larger signal improvement. This improvement comes essentially `for free', as the basis functions can be precomputed and no change in hardware or number of wavefront shaping measurements is required.

For the basis constructed from Zernike polynomials, we have improved the orthonormality by nearly 2 orders of magnitude, while preserving smoothness and symmetry in x and y.

In addition to wavefront shaping, another application of our method could be measuring subsets of the transmission matrix of an optical system.

% Conclusion
In summary, we have developed a method to orthonormalize a given set of phase-only basis functions for a given beam amplitude profile. Our method can reduce the non-orthonormality by orders of magnitude, and in special cases, yield a polynomial approximation to a perfect solution.
Furthermore, we show that the results of this method have direct impact on real-world applications, as we improve a wavefront shaping algorithm by using our orthonormalized basis as a drop-in replacement.

\bibliographystyle{unsrt}
\bibliography{main}

\end{document}

\maketitle
\hrule

\section{Experimental setup}

\begin{figure}[h]
    \centering
    \includegraphics[width=1\linewidth]{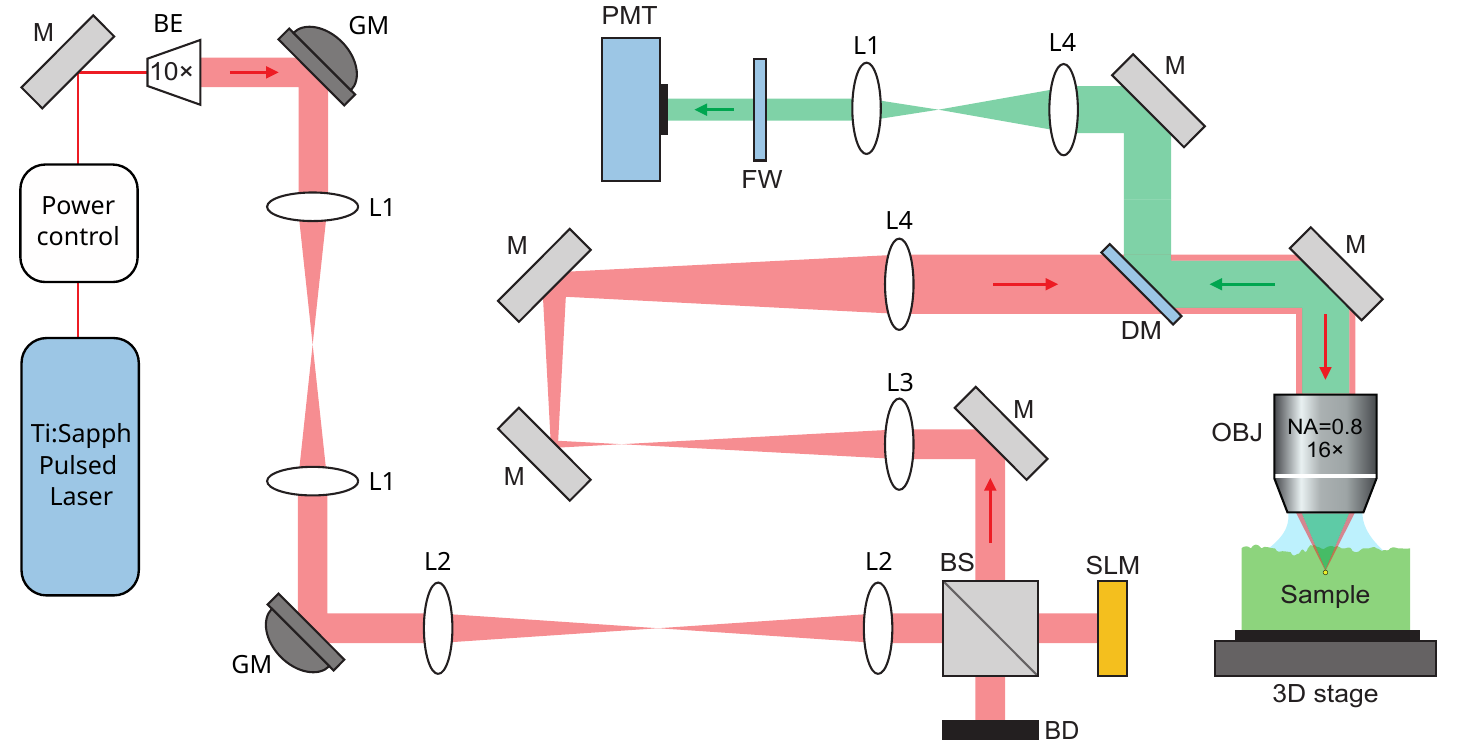}
    \caption{Schematic of the experimental setup. M: mirror, BE: beam expander, GM: galvo mirror, L: lenses with the following focal distances: L1: \SI{100}{mm}, L2: \SI{200}{mm}, L3: \SI{150}{mm} and L4: \SI{300}{mm}, BS: 50/50 beam splitter, BD: beam dump, SLM: spatial light modulator, DM: dichroic mirror, OBJ: Objective, FW: filter wheel, PMT: Photomultiplier tube. Figure adapted from \cite{thendiyammalModelbasedWavefrontShaping2020}.}
    \label{fig:2pef-microscope}
\end{figure}

Our sample is made by casting PDMS onto a 220 grit optical diffuser, causing the PDMS sample to have a surface with a controlled roughness. The PDMS has fluorescent beads dispersed inside (Polysciences Fluoresbrite, plain YG, \SI{500}{nm} microspheres).

We image the sample in our two-photon excitation fluorescence (2PEF) laser-scanning microscope (see Fig.~\ref{fig:2pef-microscope}). We use a titanium-sapphire laser (Spectra-Physics, Mai Tai) as excitation light source at a wavelength of \SI{804}{nm}. Two galvo mirrors (GM, Thorlabs GVS111/M), one for each transverse axis, are conjugated to the spatial light modulator (SLM, Meadowlark 1920x1152 XY Phase Series) with 4f-systems. The SLM is illuminated with a Gaussian amplitude profile.

The SLM is conjugated to the back pupil of the microscope objective (Nikon CFI75 LWD 16X W) with a 4f-system. On the back pupil, the Gaussian amplitude profile has a waist radius of $\rwaist=\SI{11.8}{mm}$. The back pupil has a radius of \SI{10}{mm}.

By changing the beam angle with the galvo mirrors, we can scan the focus in the sample along the transverse axes. When a fluorescent bead inside the sample is illuminated, it emits light at a shorter wavelength due to the 2PEF effect. We detect this light with a photomultiplier tube (PMT, Hamamatsu H7422-40).

We can correct aberrations from the sample's scattering surface by displaying a phase correction pattern on the SLM. See Fig.~\ref{fig:2pef-microscope} for a full schematic.

\section{Wavefront shaping with an orthonormal basis}

Although wavefront shaping works with any basis \cite{vellekoopFocusingCoherentLight2007a}, many wavefront shaping algorithms use an orthonormal or approximately orthonormal basis. In this section we will explain why orthonormal bases are optimal for wavefront shaping.

Firstly, we recall the equation linking the input and output field of an optical system through a transmission matrix \cite{kubbyWavefrontShapingBiomedical2019}:

\begin{equation}
    \label{eq:TM-sum}
    E_b = \sum_a^N t_{b a} E_a
\end{equation}
where $E_b$ indicates the output field elements, $t_{b a}$ denotes the elements of the transmission matrix, and $E_a$ indicates the input field elements. In matrix form, this becomes:
\begin{equation}
    \label{eq:TM-matrix}
    E_\mathsf{b} = T E_{\mathsf{a}}
\end{equation}
where $E_\mathsf{b}$ denotes the output field as a complex vector with $M$ elements, $E_{\mathsf{a}}$ denotes the input field as a complex vector with $N$ elements, and $T$ denotes the $M$-by-$N$ transmission matrix for the chosen input and output bases.

Secondly, if we wish to use a different basis for describing the incident field, we may express this with a change of coordinate matrix $B$. The matrix product with $T$ forms the transmission matrix $\hat{T}$ in the new basis:
\begin{equation}
    \label{eq:TM-basis}
    \hat{T} = TB
\end{equation}
In order to learn more about the implications of $B$, we'll perform a  singular value decomposition (SVD) on it:
\begin{equation}
    \label{eq:TM-SVD}
    \hat{T} = TB = T\,UDV^*
\end{equation}
where $U$ and $V$ are unitary matrices, $D$ is a diagonal matrix containing the singular values of $B$ as its diagonal elements and $^*$ indicates the conjugate transpose.
We assume, the medium is strongly scattering. For a strongly scattering medium, $T$ is a random matrix. The matrix product with the unitary matrix $U$ produces a new random matrix $\tilde{T}=TU$ with the same distribution as $T$ \cite{Beenakker1997RandommatrixTheory}. Substitution in Eq.~\ref{eq:TM-SVD} yields:
\begin{equation}
    \label{eq:new-TM-SVD}
    \hat{T} = \tilde{T} D V^*
\end{equation}

In wavefront shaping, the ultimate goal is to maximize transmission to a desired output field $E_\beta$. The incident field that achieves this is given by $E_a \propto t^*_{\beta a}$ \cite{Tanter2000TimeReversal, kubbyWavefrontShapingBiomedical2019}. Hence, we want to display $\hat{T}^*$ on the SLM to produce the output $\hat{T}\hat{T}^*$. As $V$ is a unitary matrix:
\begin{equation}
    \label{eq:TM-TMconj}
    \hat{T}\hat{T}^* = \tilde{T} D V^* \, V D^*\tilde{T}^* = \tilde{T}D \, D^*\tilde{T}^*
\end{equation}

$V^* \, V$ cancels out and therefore doesn't play a role in the end result of $\hat{T}\hat{T}^*$. Thus, we only need to consider the diagonal matrix $D$ to understand the effect of $B$ on the expected performance of our wavefront shaping experiment. In this regard, $D$ effectively acts as an amplitude mask for the input of $\tilde{T}$: each input field for $\tilde{T}$ gains a factor $D_a$. We can compute the influence of this effective amplitude factor in the same manner we compute the influence of regular amplitude factors $A_a$. For a large number of input modes $N$ \cite{Akbulut2011FocusingLight, Mastiani2024PracticalConsiderationsb}:
\begin{equation}
    \label{eq:gamma-Aa}
    \left\langle |\gamma_A|^2 \right\rangle = \frac{\overline{A_a}^2}{\overline{A_a^2}}
\end{equation}
where $\left\langle |\gamma_A|^2 \right\rangle$ is the expected fidelity reduction factor due to the amplitude distribution and $A_a$ is the amplitude factor of basis function $a$. The overline indicates averaging over all modes $a$. From the Cauchy-Schwarz inequality it follows that $\overline{A_a}^2 \leq \overline{A_a^2}$ and so $\left\langle |\gamma_A|^2 \right\rangle \leq 1$.
Similarly:
\begin{equation}
    \label{eq:gamma-Da}
    \left\langle |\gamma_B|^2 \right\rangle = \frac{\overline{D_{aa}}^2}{\overline{D_{aa}^2}}
\end{equation}
where $\left\langle |\gamma_B|^2 \right\rangle$ is the expected fidelity reduction factor due to the choice of basis and $D_{aa}$ are the diagonal elements of the SVD of the change of coordinate matrix $B$ for that basis. Lastly, we would like to note the special relation between $D$ and the Gram matrix $G$. Each element of the Gram matrix is an inner product of a pair of basis vectors.
\begin{equation}
    \label{eq:gram-matrix-form}
    G = B^* B
\end{equation}
Inserting the SVD of $B$ yields:
\begin{equation}
    \label{eq:gram-matrix-form}
    G = U^*D^*V \, V^*DU = U^*D^* \, DU = U^* D^2 U
\end{equation}
Therefore, $D_{aa}^2$ are the eigenvalues of the Gram matrix $G$.
For an orthonormal basis, $G$ is the identity matrix and $D_{aa}^2=D_{aa}=1$ for all $a$. From Eq.~\ref{eq:gamma-Da} it follows that $\left\langle |\gamma_B|^2 \right\rangle$ then has the optimal value of 1. Therefore, an orthonormal basis is the optimal choice of basis.

\section{Special case solution for the coordinate transform}
\label{sec:special-case-plane-wave}

In this section, we will explain why orthonormalizing the plane wave basis from Mastiani~et~al. \cite{mastianiWavefrontShapingForward2022} is a special case of the problem, which has a perfect solution. We will show that our method has found polynomial approximations to this solution.

Consider the plane wave basis $\{\fun_{\im}\}$ with a constant amplitude:
\begin{equation}\label{eq:basis-function-pw}
    \fun_{\im}(x, y)
    = \amp_R(x,y)\, e^{i\phi_m(x, y)}
    = \amp_R(x,y)\, e^{i \kx x + \ky y}
\end{equation}
where $\im$ denotes the basis function index, $\phasexy{\im}$ is real and describes the phase, and $\kx$ are integer multiples of $2\pi$ and $\ky$ are integer multiples $\pi$. $\amp_R(x,y)$ is the following rectangularly shaped amplitude function:
\begin{equation}\label{eq:rectangular-amplitude}
    \amp_R(x,y) =
    \begin{cases}
        \frac{1}{\sqrt{2}}
        &\text{if }x\in [-1, 0] \text{ and } y\in [-1, 1] \\
        0  & \text{otherwise}
    \end{cases}
\end{equation}
These values make $\{\fun_{\im}\}$ orthonormal:
\begin{equation}\label{eq:inner-product-pw}
    \inner{\fun_{\im}}{\fun_{\imm}}
    = \iint_{-\infty}^{\infty}
    \amp^2_R(x,y) \,
    e^{i(\phasexy{\imm}-\phasexy{\im})} 
    \;dx\,dy = I_{\im\imm}
\end{equation}
where $I$ denotes the identity matrix.

However, the amplitude profile in our setup is not a rectangular shape, but defined by the beam amplitude profile $\ampxy$. $\ampxy$ is real and non-negative, and $0$ where $x\rightarrow\pm\infty$ or $y\rightarrow\pm\infty$. As described in the paper, we transform the coordinates for $\phasexy{\im}$ to get an alternative phase function $\wphasexy{\im}$. These two things give us the new basis function $\wfun_{\im}$:
\begin{equation}\label{eq:basis-function-new}
    \wfun_{\im}(x,y) = \ampxy \,e^{i\phi_\im(\transf{x}, \transf{y})}
\end{equation}
The corresponding inner products are:
\begin{equation}\label{eq:inner-product-new}
    \inner{\wfun_{\im}}{\wfun_{\imm}}
    = \iint_{-\infty}^{\infty}
    \amp^2(x,y) \,
    e^{i(\wphasexy{\imm}-\wphasexy{\im})} 
    \;dx\,dy
\end{equation}
Now let's consider that we choose a coordinate transformation whose Jacobian $J$ \cite{Stewart2008CalculusEarly}
\begin{equation}\label{eq:jacobian-j}
    J = 
    \frac{\partial\transf{x}}{\partial x}
    \frac{\partial\transf{y}}{\partial y} -
    \frac{\partial\transf{x}}{\partial y}
    \frac{\partial\transf{y}}{\partial x}
\end{equation}
satisfies the following requirement:
\begin{equation}\label{eq:jacobian-requirement}
    \amp^2_R(x',y') \abs{J} = \amp^2(x,y)
\end{equation}
If the transformation is one-to-one within the area where $A(x,y)>0$ then we may apply the following substitution \cite{Stewart2008CalculusEarly}:
\begin{equation}\label{eq:dxdy}
    \amp^2(x,y) \;dx\,dy =
    \amp^2_R(\transf{x},\transf{y}) \;d\transf{x}\,d\transf{y}
\end{equation}

We use Eq.~\ref{eq:dxdy} to do a change of coordinates for the integral of Eq.~\ref{eq:inner-product-new}:
\begin{equation}\label{eq:inner-product-substitute}
    \inner{\wfun_{\im}}{\wfun_{\imm}}
    = \iint_{-\infty}^{\infty}
    \amp^2_R(\transf{x},\transf{y}) \,
    e^{i(\wphasexy{\imm}-\wphasexy{\im})} 
    \;d\transf{x}\,d\transf{y}
\end{equation}

This integral is equivalent to Eq.~\ref{eq:inner-product-pw} and so $\inner{\wfun_{\im}}{\wfun_{\imm}}=I_{\im\imm}$. Thus, when the coordinate transform satisfies the differential equation from Eq.~\ref{eq:jacobian-requirement}, it restores perfect orthonormality for the new amplitude profile $\ampxy$.

\begin{figure}
    \centering
    \includegraphics[width=1\linewidth]{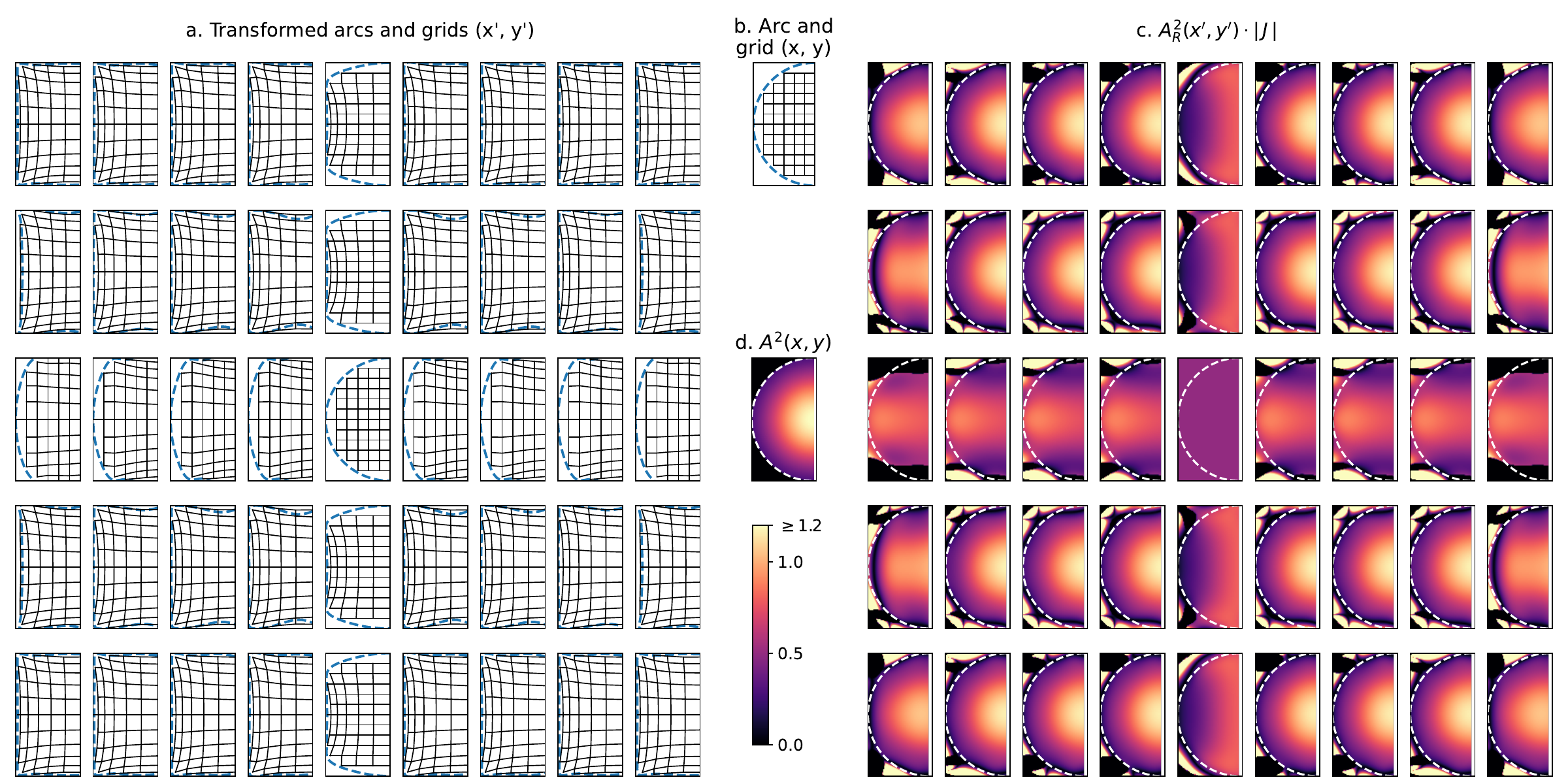}
    \caption{
    Visualizations of the transformation of each basis function for the plane wave basis. (a) shows the domain of transformed coordinates $\transf{x}\in[-1, 0]$ (horizontal) and $\transf{y}\in[-1, 1]$ (vertical), where each subplot has its own transform.
    Blue dashed curve: a curve with transformed coordinates $(\transf{x}, \transf{y})$, for which the original coordinates $(x,y)$ form an arc with radius $1$.
    Black solid curves: transformed grid formed with transformed grid points $(\transf{x}, \transf{y})$. Their original grid points $(x,y)$ lie within a radius of $1$. (b) shows a non-transformed version of the arc and grid. (b, c, d) show the domain $x\in[-1, 0]$ (horizontal) and $y\in[-1, 1]$ (vertical).
    (c) For each transformation, $\amp_R^2(\transf{x}, \transf{y})\abs{J}$ is shown, where $\amp_R(x, y)$ is the rectangularly shaped amplitude function.
    (d) The amplitude profile squared $\amp^2(x,y)$. In (c, d), the white dashed line indicates the objective's pupil aperture. Outside this pupil aperture, $\ampxy=0$.}
    \label{fig:grids-jacobian}
\end{figure}

With this insight, we look at the coordinate transforms found by our method for orthonormalizing the plane wave basis from Mastiani~et~al.~ \cite{mastianiWavefrontShapingForward2022}.
Fig.~\ref{fig:grids-jacobian}b shows a square grid (solid black) within the pupil aperture and an arc (dashed blue) of radius $1$ indicating the pupil aperture. Outside the pupil aperture, $\ampxy=0$.
Fig.~\ref{fig:grids-jacobian}a shows the same grid and arc but with transformed coordinates $\transf{x}(x,y), \transf{y}(x,y)$.
We observe that most transformations deform the arc to form a path along the rectangular discontinuous boundary of $\amp_R(x,y)$.
In other words, the `boundary' of $\ampxy$ is transformed to match the `boundary' of $\amp_R(x,y)$. An exception are the basis functions in the middle row and middle column. These basis functions have no phase variation in $x$ and/or $y$. We observe that their corresponding transformed grids remain unchanged in $x$ and/or $y$.

Next, we will test if the transformations found by our method satisfy the requirement of Eq.~\ref{eq:jacobian-requirement}.
As the transformation found by our method $\transf{x}(x,y), \transf{y}(x,y)$ are two bivariate polynomials, $J$ can be computed according to the basic derivative rules for polynomials. With this, we have computed both the left-hand side (Fig.~\ref{fig:grids-jacobian}c) and right-hand side (Fig.~\ref{fig:grids-jacobian}) of Eq.~\ref{eq:jacobian-requirement} for each transformation.
As can be observed, within the pupil (white dashed line), most plots of $\amp_R^2(\transf{x}, \transf{y})\abs{J}$ are very similar to the plot of $\amp^2(x,y)$. This strong similarity means that Eq.~\ref{eq:jacobian-requirement} approximately holds for the transformations found by our method. In other words, our method has found polynomial approximations to a transformation that restores perfect orthonormality. Exceptions are again the plots in the middle row and middle column. These basis functions have no phase variation in $x$ and/or $y$.

In conclusion, in the special case when the initial basis is orthonormal for a different amplitude profile, there exists an ideal coordinate transformation that achieves perfect orthonormality for a desired amplitude profile. The plane wave basis is such a special case, and our method has found polynomial approximations to this ideal coordinate transformation.

\section{Iterative dual reference wavefront shaping}

We use the dual reference wavefront shaping algorithm \cite{mastianiWavefrontShapingForward2022} with a 2PEF feedback signal. Using a 2PEF feedback signal from an aberrated focus means that the signal is not well localized, which prevents selecting a single location as target.
Still, it is possible to converge to a single localized target and form a sharp focus by iterative wavefront shaping \cite{Osnabrugge2019BlindFocusing}.
The downside is that this limits the algorithm to measuring one target at the same time. In other words, we measure a one-row transmission matrix, which corresponds to one correction pattern for our one target location in the sample.

We implement an iterative version of the dual reference algorithm. In short, we perform the following steps in our measurement procedure:
\begin{enumerate}
    \item \label{item:prescan} We record an image with the laser scanning microscope. Due to aberrations, this image is blurry, but it is still possible to see the approximate locations of features in the image.
    \item We select the brightest feature in the image. We adjust the scanner settings to zoom to the region of interest (ROI), with the feature centered and the size of the ROI on the order of the diffraction limit.
    \item We divide the SLM in two parts of equal size. We'll call them side A and side B.
    \item We use side A to display the phase patterns and produce the orthonormal fields.
    \item We keep side B flat and use it as a reference field.
    \item \label{item:phasestep} We display several phase-stepped versions of each phase pattern on side A while keeping side B static.
    \item \label{item:record} With each phase step, we record the intensity inside the region of interest. This allows us the reconstruct the relation between the input phase pattern and output field (i.e. the corresponding transmission matrix element) using phase stepping interferometry \cite{kubbyWavefrontShapingBiomedical2019}.
    \item When all phase patterns have been displayed, we have measured the transmission matrix (which has one row) for this side of the SLM.
    \item The transmission matrix is defined in our chosen basis. In order to create a usable correction pattern, we transform the transmission matrix row to SLM coordinates and conjugate it.
    \item \label{item:switch} We repeat steps \ref{item:phasestep}-\ref{item:record} but with two changes: side A and B switch roles, and we use the newly computed transmission matrix to display a correction pattern on the reference side; a `shaped reference'.
    \item We repeat steps \ref{item:phasestep}-\ref{item:switch} two more times, but now we use a shaped reference every time for both sides. We do 6 iterations in total (3 per side).
    \item Finally, we compute the transmission matrix for the entire SLM, transform it to SLM coordinates, and display it on the SLM.
    \item \label{item:finalscan} We set the scanner to image a large field-of-view and record the final image.
    \item We move the sample with a motorized stage to bring a new sample location in the field of view. We repeat steps \ref{item:prescan}-\ref{item:finalscan} at the new location in the sample. We have measured ?? %%% TODO: how many locations?
    locations in total in the same sample.
\end{enumerate}

For further details on the wavefront shaping algorithm, please see \cite{vellekoop2023Openwfs}.
%%%% TODO: waar doen we het meetscript? In OpenWFS of in mijn phase-only-orthonormalization repo?

\bibliographystyle{unsrt}
\bibliography{supplementary}